\documentclass[12pt]{iopart}
\usepackage{iopams}
\usepackage{epsfig}
\begin{document}

\title{Relaxed constraints on neutrino oscillation parameters}

\author{Daniela Kirilova\dag \ddag \footnote[3]{To
whom correspondence should be addressed (dani@astro.bas.bg)}
\ and Mariana Panayotova\dag}

\address{\dag\ Institute of Astronomy,  Sofia, Bulgaria}

\address{\ddag\ Abdus Salam ICTP, Trieste, Italy}

\begin{abstract}
We study the cosmological  constraints on active-sterile neutrino
oscillations $\nu_e \leftrightarrow \nu_s$ for the case when $\nu_s$ is
 partially filled initially, i.e. $0 < \delta N_s < 1$.
We provide numerical analysis of the cosmological production of $^4\!$He, in
the
presence of  $\nu_e \leftrightarrow \nu_s$ oscillations, effective
after neutrino decoupling, accounting for all known
oscillations effects
on cosmological nucleosynthesis.
Cosmological constraints on
oscillation parameters
 corresponding to  higher than $5\%$ $^4\!$He 
overproduction and different non-zero initial populations of the
sterile state  $\delta N_s<1$ are calculated. These generalized cosmological constraints corresponding to
$\delta N_s >0$ are
 relaxed in comparison with the
$\delta N_s=0$ case and the relaxation is proportional to $\delta N_s$.


\end{abstract}



\maketitle

\section{Introduction}
Atmospheric, solar
and  terrestrial neutrino oscillations experiments provided evidence for
neutrino oscillations.
The atmospheric neutrino data analysis confirmed the $\nu_\mu \leftrightarrow \nu_\tau$ channel
as the dominant atmospheric oscillation solution. The results of the terrestrial experiment KamLAND and the results
of SNO salt
phase prefer flavor oscillation LMA solutions to the $\nu_e \leftrightarrow \nu_s$ one for the solar neutrino problem.
Active-sterile
neutrino oscillations are discussed as a supplementary sub-dominant channel. Even after
 the confirmation
of LMA solution as a dominant solution of the solar neutrino problem, the oscillation parameters are
not precisely known - the completeness of LMA solution was questioned and the scope for some
possible sub-dominant transitions was explored \cite{al,kaki2,kaki1,chapu,hosmir,desan1,desan2,pulchau}. 
The available results of the neutrino
oscillations from the solar experiments,
including the later SNO ES, CC and NC data, slightly favor the existence of
a small sterile oscillating sector. Thus, it is still interesting to explore the cosmological
influence of sterile oscillating neutrinos.

Neutrino $\nu_e \leftrightarrow \nu_s$ oscillations affect the expansion rate and the neutrino involved processes in the early
Universe and in particular, influence Big Bang Nucleosynthesis (BBN) \cite{dol2,ki5}.
This allows to put  stringent constraints
on neutrino $\nu_e \leftrightarrow \nu_s$ oscillations  from
BBN considerations. The whole LMA and partially LOW $\nu_e \leftrightarrow \nu_s$ solar
oscillation solutions and atmospheric $\nu_e \leftrightarrow \nu_s$ solution 
were excluded by cosmological considerations \cite{bado2,bado1,enq,chiki,ki1,kichi3,kichi2,dol1,di,cir,ciret,dolvil,dol5,kichi5}.

Detail reviews discussing neutrino role in cosmology and neutrino masses and mixings may be found in refs. \cite{dol5,kichi5,stru,lesg}.

Cosmological constraints were obtained usually assuming zero sterile neutrino
state population before neutrino oscillations epoch, $\delta
N_s=\rho_{\nu_s}/\rho_{\nu_e}^{eq}=0$, where $\rho_{\nu_e}^{eq}$ is the equilibrium electron
neutrino energy density. However, sterile neutrinos production is predicted by different
types of models, like GUT models, models with large extra dimensions, many-fold Universe models, mirror matter models, 
neutrino oscillations models, etc., and 
may be 
present at the onset of BBN epoch. Hence, the degree of population
of $\nu_s$ and its initial energy spectrum distribution
depends  on the $\nu_s$ production model, and in general $\delta
N_s \ge 0$.
% \cite{gohmoh}.

The general case of initially (before oscillations) non-zero sterile neutrino
population was examined and its influence on the oscillation effects at BBN epoch was
discussed in reference \cite{ki2}. It was found that {\it the kinetic effect
of oscillations is sensitive to the initial
population of $\nu_s$}. BBN constraints
on neutrino oscillations parameters
in the general case of  sterile neutrino state,
partially occupied initially, were discussed  in reference \cite{ki3}
for oscillations effective after electron neutrino decoupling.
The  case of $3\%$ $^4\!$He overproduction was  considered.
 {\it The presence of  non-empty  sterile neutrino $0<\delta N_s<0.54$ before oscillations
was shown to lead to strengthening of the cosmological constraints
proportional to the  $\delta N_s$ value}.

Recent studies of the systematic uncertainties of primordial $^4\!$He abundance
suggest as a more reliable value of  $^4\!$He observational uncertainty  $\delta Y_p/Y_p>5\%$.
In what follows we derive 5.2$\%$ cosmological constraints, corresponding to $\delta Y_p/Y_p>5\%$ $^4\!$He overproduction on
$\nu_e \leftrightarrow \nu_s$ oscillation parameters for the same type of non-equilibrium
neutrino oscillations. We show that in contrast to $3\%$ $^4\!$He constraints,
 {\it in the $\delta Y_p/Y_p>5\%$ case the non-zero initial sterile population leads to a
relaxation of the constraints on oscillation parameters proportional to the value of $\delta N_s$},
 and finally to their alleviation  for  $\delta N_s=1$.

In the next section, we discuss briefly the primordial $^4\!$He production
in the case of  non-zero  sterile population at the start of oscillations.
In the last section we present the cosmological constraints
corresponding to $^4\!$He-overproduction $\delta Y_p/Y_p>5\%$ and different levels of
initial population of the sterile state  and compare them
to $\delta N_s=0$ case.

\section{Production of primordial $^4\!$He and neutrino oscillations}

\subsection{Primordial $^4\!$He}
\label{int}
$^4\!$He is the most abundantly produced, most precisely measured
and calculated element among the primordially formed light elements.
  It has also a
simple post-BBN chemical
evolution - it is only produced in stars. 
 Therefore, it is the preferred light element used for
obtaining limits on nonstandard physics.
 Primordially produced $^4\!$He is calculated with
great precision \cite{lotur,cybet2,cyb,cu,co} $Y_p = 0.2485\pm0.0005$.
The predicted $^4\!$He value is in
agreement with the observational data for $^4\!$He inferred from astrophysical
observations  \cite{ol,izthu2,izthu1,olskil,fisar,cybet1,fuku}.

Currently there are significantly different observational determinations of the
primordial
$^4\!$He abundance, which point to the existence of greater systematic errors than assumed before.
Izotov and Thuan have found $Y_p \approx 0.242\pm0.009$ for 82 HII regions and 
$Y_p = 0.2421\pm0.0021$ for 7 preferred HII regions, and $Y_p = 0.2429\pm0.0009$ for an extended 
data set of 89 HII regions \cite{izthu2,izthu1}. 
Olive and Skillman have re-analyzed their data and gave $Y_p\approx 0.245\pm0.013$ 
for all 82 HII 
regions and $Y_p = 0.2491\pm0.0091$ for the 7 preferred targets \cite{olskil}. 
The real uncertainty on the primordial $^4\!$He abundance may be larger than previously assumed $3\%$. 
Present determinations of primordial $^4\!$He abundance
from observations of extragalactic HII regions,
indicate a significantly grater uncertainty for the $^4\!$He mass fraction 
\cite{izthu1,olskil,cybet1,fuku} namely $dY/Y\sim 5\%$. \footnote{See also the pioneer papers discussing the possibility for higher 
systematic errors in the determination of primordial $^4\!$He abundance, which was believed to be much smaller 
than at present
\cite{sasgold,lur,lur1}.}

Thus a derivation of cosmological constraints on
oscillation parameters corresponding to about $5\%$ $^4\!$He overproduction seems relevant.

The  primordial  $^4\!$He  abundance
 essentially
depends  on the freeze-out of nucleons, which occurs when in the process of expansion the
weak processes rates $\Gamma_w$,  governing the
neutron to proton transitions,
become comparable to the expansion rate  $H(t)$.
So, the primordially
produced mass fraction of $^4\!$He
$Y_p \sim 2(n/p)_f/(1+n/p)_f$,  
is a strong function of relativistic degrees of freedom at BBN epoch, 
which enter through $H(t)$, where $H(t)\sim\sqrt{G_N g_{eff}}~T^2$.
\footnote[3]{Due to its strong dependence on $g_{eff}$, $^4\!$He abundance was  
used to put constraints on the number of the relativistic particles during
BBN epoch \cite{shva,steig,li}, parametrized by $\delta N_{\nu}=N_{\nu}-3$. 
Cosmological data (except BBN) from CMB, LSS, SN, are not very restrictive on the extra 
light particles \cite{trotta,hann,hann1,cirre}.} $Y_p$  also depends on the electron 
neutrino characteristics, namely  neutrino energy spectrum, 
number densities $N_{\nu}$  
and  the  neutrino-antineutrino asymmetry, entering through  $\Gamma_w$,
 $\Gamma_w \sim G_F^2 E_{\nu}^2 N_{\nu}$.

\subsection{Oscillations effects}

Non-empty initial sterile neutrino  population, present at the start of oscillations, influences  
 BBN by (i) increasing the expansion rate and
(ii) suppressing  the kinetic
effects of $\nu_e\leftrightarrow \nu_s$ on BBN.

\begin{enumerate}
\item {\it The presence of} $\delta N_s$ means increased  effective number of neutrino species in equilibrium
during BBN, and
  {\it leads to faster expansion}
of the Universe, $H(t)\sim g^{1/2}_{eff}$  causing earlier $n/p$-freezing, $T_f\sim (g_{eff})^{1/6}$, hence,  
 $^4\!$He {\it is overproduced}. This dynamical effect is parametrized through $\delta Y_d \sim 0.013 \delta 
N_s$.

\item In case of oscillations between
non-equilibrium sterile neutrino state $0< \delta N_s<1$ and
electron neutrino, proceeding  after $\nu$ decoupling,  $\nu_e \leftrightarrow \nu_s$ oscillations lead
  to considerable
and continuous deviations from the equilibrium $\nu_e$ spectrum
(spectrum distortion) and production of neutrino-antineutrino asymmetry.\footnote{Neutrino-antineutrino asymmetry
is  generated during the $\nu_e \leftrightarrow \nu_s$
resonant oscillations \cite{chiki}.
This dynamically produced asymmetry suppresses
oscillations at small mixing angles. Therefore, it leads to
underproduction of $^4\!$He and alleviates BBN
constraints on oscillation parameters compared to the case without the account of
asymmetry growth. We have accounted in this work for its effect as well, however the effect is
sub-dominant and will not be discussed further on.}

 This effects nucleons kinetics
in the pre-BBN epoch \cite{ki5,ki4,ki2}.
This kinetic effect may be parametrized in terms of additional neutrino $\delta N_{kin}$ and will be 
denoted further $\delta Y_{kin} \sim 0.013 \delta N_{kin}$. {\it Non-zero}  $\delta N_s$ {\it suppresses the 
kinetic effect 
of oscillations}:  $\delta N_{kin}$ decreases with the increase of $\delta N_s$, hence $\delta N_s$ 
{\it leads to a decrease of the overproduction of} $^4\!$He.

\end{enumerate}

 For a wide range of oscillation
parameters the kinetic effect of oscillations  is large during the period of freezing of the nucleons
and, therefore, affects BBN. This effect  is strong  even when there is a  considerable 
population of the sterile neutrino
state  at the onset of the electron--sterile oscillations. 
The $\nu_e$ energy spectrum distortion is the greatest, if the sterile state is
empty at the start of oscillations, $\delta N_s=0$ and  decreases
with the increase of the sterile state degree of population \cite{ki2}. 
The same behavior is to be expected for the kinetic effect of oscillations. 

The total effect of (i) and (ii) can be approximately 
 described by $\delta Y_p \sim 0.013 \delta N$ $\delta N=\delta N_s + \delta N_{kin}$, 
where $\delta N_{kin}=\delta N_{kin}^{max}(1-\delta N_s)$ and
$\delta N_{kin}^{max}$ is the kinetic oscillations effect, corresponding to
 $\delta N_s=0$.
 The expression presents a   good approximation to the numerically calculated
   dependence of the kinetic effect on the initial population of $\nu_s$,
 derived  in reference \cite{ki2}:

There is an interesting interplay between the different effects
which  $\delta N_s\ne 0$ exerts on oscillations and on BBN with oscillations
and hence, the $\delta Y_p$ production and the cosmological constraints on oscillations parameters for the
case
$\delta N_s \ne 0$  differ from the ones derived in references \cite{kichi3,kichi2,kichi1} for  $\delta N_s=0$.

As found in  \cite{ki3}, for $\delta Y_p/Y_p>5\%$ corresponding to $\delta N_{kin}^{max}>1$, 
the suppression effect (ii) dominates over the dynamical effect (i) of  $\delta N_s\ne 0$. Hence, the total effect 
is a decreasing function of 
$\delta N_s$, i.e. $^4\!$He overproduction decreases with $\delta N_s$ (in comparison with the case $\delta N_s=0$) and 
correspondingly the BBN constraints on oscillation parameters relax.
In the opposite case $\delta Y_p/Y_p<5\%$,  corresponding to $\delta N_{kin}^{max}<1$, the dynamical effect (i) dominates 
and the total effect is increasing 
with $\delta N_s$. I.e. $^4\!$He overproduction increases and the BBN constraints on oscillations strengthen 
in comparison with the case $\delta N_s=0$.  In the case $\delta Y_p/Y_p=5\%$ the constraints for $\delta N_s\ne0$  
coincide with the ones 
 for  $\delta N_s=0$, due to the cancellation of the two effects (i) and (ii). 

For illustration of $\delta N_{kin}^{max}>1$ case, which we consider further on, we present in Figure 1 the 
effects (i) 
and (ii) on $^4\!$He overproduction 
at $\delta m =10^{-7}$ eV$^2$ and $\sin^2 2\theta = 1$. 
We have studied numerically the contribution of these effects on neutrons to nucleons freezing ratio 
$X_n=n_n^f /n_{nuc}$ for different
$\delta N_s$. The primordial yield of helium to a good approximation is expressed through  $X_n$: 
$Y_p\sim X_n \exp(-t/ \tau_n)$, where $\tau_n$ is the neutron lifetime.
 For the chosen set of parameters the $^4\!$He overproduction
decreases with the increase of $\delta N_s$. The suppression
effect (ii) of $\delta N_s$ dominates and a relaxation of the cosmological constraints 
compared to $\delta N_s=0$ must be expected.

\begin{figure}
\centerline{\epsfxsize=12cm \epsfysize=8cm \epsfbox{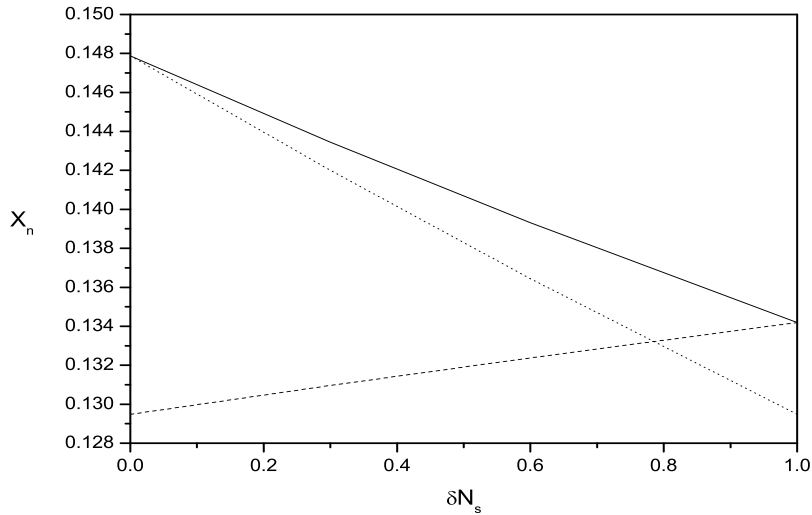}}

\caption{The solid curve presents the frozen neutron density relative to nucleons $X_n=n_n^f/n_{nuc}$ as a 
function of the sterile neutrino initial population, at $\delta m =10^{-7}$ eV$^2$ and $sin^2 2\theta = 1$.
The dotted curve presents the kinetic effect, while the dashed curve presents energy density increase effect.}.
\end{figure}

In the next section we present and discuss 
the calculated cosmological constraints on oscillation parameters corresponding 
to higher than $5\%$ $^4\!$He overproduction and  $0<\delta N_s<1$.

\section{Cosmological constraints on oscillation
parameters}

BBN constraints corresponding to  $\delta Y_p/Y_p=3\%$ overproduction of
 $^4\!$He and non-zero initial population of the sterile neutrino
 $\delta N_s<0.54$ were calculated recently \cite{ki3}. As far as $\delta Y_p/Y_p=3\%$ corresponds to 
$\delta N_{kin}^{max} <1$, the
constraints strengthen with the  increase of the $\delta N_s$ value. 
They increase the BBN $3\%$ $^4\!$He exclusion region for oscillation parameters corresponding to $\delta N_s=0$ towards
smaller  $\delta m^2$. 

Having in mind the existence of a large systematic 
error of $^4\!$He measurements, 
in this work we provide a numerical calculation of cosmological constraints
corresponding to  $\delta Y_p/Y_p > 5\%$
$^4\!$He overproduction ($\delta N_{kin}^{max}>1$) and different initial degrees of sterile neutrino 
population in the range $0\le\delta N_s<1$. We have chosen   
$\delta Y_p/Y_p=5.2\%$, i.e. a value slightly higher than the critical one $5\%$, inorder to
illustrate the possibility for relaxation of the cosmological constraints on oscillations for $\delta N_s\ne 0$.
 
Our numerical analysis has shown that cosmological
constraints corresponding to $5.2\%$ $^4\!$He overproduction  relax with the increase of $\delta N_s$ values. 
Up to  $\delta N_s=0.5$ the cosmological
constraints   are slightly
 relaxed in comparison with the case $\delta N_s=0$,  however, for higher $\delta N_s$ values,
the constraints relax noticeably. 
The reason for that is the predominance of (ii) effect  
(the suppression of the oscillations kinetic effects) over the dynamical effect (i) for the given uncertainty 
of $^4\!$He. All cosmological constraints
 corresponding to $\delta Y_p/Y_p>5\%$ will have such behaviour, namely, they will be relaxed in comparison to the   
constraints for $\delta N_s = 0$. (Vice versa, the constraints corresponding to lower than $5\%$ $^4\!$He uncertainty
will be more stringent than the ones for  initially empty sterile neutrino state.)

\begin{figure}
\centerline{\epsfxsize=12cm \epsfysize=8cm \epsfbox{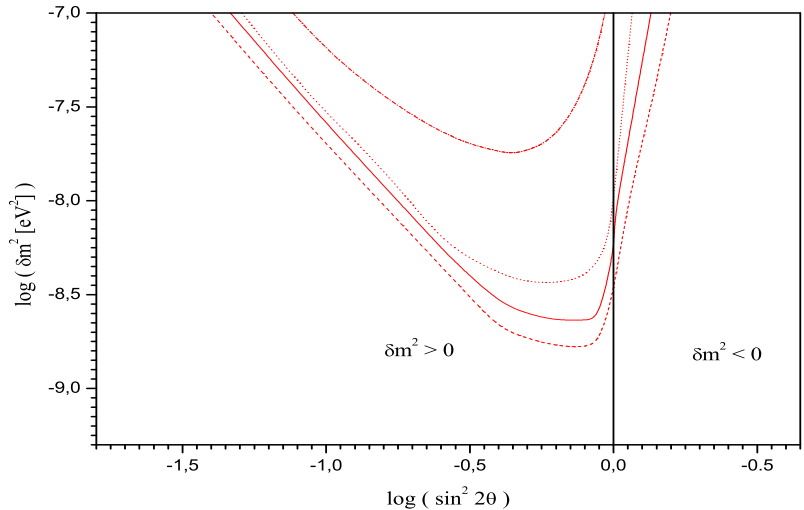}}
%\epsfbox{dy005.eps}
%\includegraphics[width=12cm,height=8cm]{marY05.EPS}
%\includegraphics[width=12cm,height=8cm]{dy005.eps}

\caption{The dashed contour presents $\delta Y_p/Y_p=5.2\%$  BBN constraints for
$\delta N_s=0$, the solid curve corresponds to  $\delta N_s=0.5$, the dotted and the dot dashed
contours --- $\delta N_s=0.7$ and
to $\delta N_s=0.9$, respectively. The resonant oscillations case corresponds to l.h.s of 
the figure, the non-resonant one to the r.h.s..}
\end{figure}

In Figure 2 $\delta Y_p/Y_p =5.2\%$  BBN constraints are presented for different values of the
initial population of the sterile state, namely the lowest dashed contour corresponds to a
zero population, the solid curve corresponds to  $\delta N_s=0.5$, the dotted and the dot dashed
contours to  $\delta N_s=0.7$ and to $\delta N_s=0.9$, respectively.

 Another interesting result is that there are considerable  
constraints even for a very high $\delta N_s$ values for that really high $^4\!$He uncertainty. I.e. the constraints
 on neutrino mixing parameter  vanish  only when the sterile state is in equilibrium before oscillations, when  the
kinetic effect due to neutrino spectrum distortion disappears, i.e.  $\delta N_s=1$.

\section{Conclusions}
We have studied BBN  constraints on neutrino
$\nu_e \leftrightarrow \nu_s$ oscillations for the
specific case when the sterile neutrino is partially filled initially $0<\delta N_s<1$.
Non-zero  $\delta N_s$ has two-fold effect on BBN with neutrino oscillations: a dynamical effect leading to
overproduction of He-4  and a kinetic effect, leading to underproduction of He-4
 in comparison with the case of $delta N_s=0$.
So, depending on the interplay between these opposite effects, the cosmological constraints may be
  either relaxed or strengthened.

We have provided  detail numerical analysis of the BBN production of $^4\!$He,
 $Y_p$, in the presence of $\nu_e \leftrightarrow \nu_s$ neutrino oscillations,
 effective after electron neutrino decoupling.
We have calculated  and discussed  cosmological constraints on oscillation
parameters, corresponding to higher than $5\%$ uncertainty of helium-4, for non-zero initial population of
the sterile state $\delta N_s<1$.
 It was found that the cosmological constraints on oscillation parameters relax with the increase of $\delta N_s$.
 The cosmological constraints for the cases
$\delta N_s \le
0.5$ are slightly
changed
in comparison to $\delta N_s=0$ case,  
however, for bigger  $\delta N_s$ the constraints are relaxed
considerably and for  $\delta N_s=1$
 they are alleviated.

Resuming the results of the works discussing  $\delta N_s$ effect on BBN: Cosmological constraints corresponding to
higher than $5\%$ uncertainty of helium-4,  relax with the
increase of the  initial
population of
the sterile state, while the constraints corresponding to  lower than $5\%$ uncertainty of helium-4 strengthen with
$\delta N_s$.

It is remarkable, that in case of BBN with non-equilibrium oscillations between electron and sterile
neutrinos, it is possible to obtain cosmological constraints on oscillation parameters even in the  
case when the $^4\!$He abundance is known with uncertainty greater than 5$\%$ (Actually, it is possible to derive
constraints on neutrino oscillation parameters for He-4 uncertainty up to 32$\%$ in the resonant oscillations case, and
up to 14$\%$ uncertainty in the nonresonant oscillations case, as far as these are the maximal possible
helium overproduction values \cite{Ap}).
 The  cosmological constraints  persist  while   initially the sterile state is non equilibrium.
When $\delta N_s=1$ initially, the kinetic effect of oscillations is zero and
it is not possible to obtain constraints on oscillation parameters in the discussed  BBN models with non-equilibrium
electron-sterile neutrino oscillations, as well as in BBN models with equilibrium
electron-sterile neutrino oscillations.

\ack {The authors thank the referee for the useful comments and suggestions.
D. Kirilova appreciates  the visiting position at ULB, Bruxelles and the  
Regular  Associateship  of  the Abdus Salam ICTP, Trieste.  
This work  was supported in part by Belgian Federal Government
(Federal Public Planning Service Science Policy (PPS Science Policy)).}

\section{Referencing}

\end{document}